# A Z' interpretation of Bd->K*μμ data and consequences for high energy colliders

François Richard

Laboratoire de l'Accélérateur Linéaire, IN2P2–CNRS et Université de Paris–Sud, Bât. 200, BP 34, F–91898 Orsay Cedex, France.

**Abstract:** In this note I examine the possible consequences for high energy colliders of a Z' interpretation of the LHCb anomaly observed in the K*μμ final state. Two examples are elaborated in the framework of the so-called 331 model. In the first one it is shown that LEP2 provides the tightest lower mass limit for the Z' boson, above 8 TeV, while in the second one the lower mass limit is set by ATLAS/CMS to about 3 TeV. It is then shown that precision measurements at ILC 500 GeV can fully explore the underlying structure of the model by measuring the fermion final states separately: leptons, charm, beauty and top final states. Z'-Z mixing can also be substantial, thus leading to possible effects almost observable at LEP1 and which can be precisely measured at GigaZ. Discovery prospects for heavy bosons and heavy fermions at LHC are also discussed .

## I.    Introduction

Recently the LHCb experiment [1] has shown some evidence for a significant deviation from the SM in a detailed angular analysis of the channel K*μμ (see figure 1). This effect requires confirmation using the full data available sample (3 times more statistics) and a thorough estimate of the theoretical uncertainties on the SM predictions.

A theoretical interpretation [2] has been suggested in terms of a Z' boson which has a vector coupling to μμ and has a tree-level coupling to bs. This explanation is funded on an extension of the SM symmetry from $SU(2)_L$ to $SU(3)_L$, the so-called '331 model' with a $SU(3)_c \otimes SU(3)_L \otimes U(1)_X$ symmetry group.

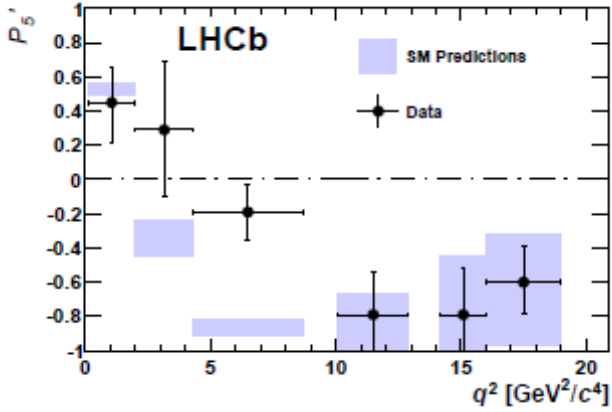

Figure 1: The observable known as P'5 is shown as function of the invariant-mass squared of μμ.

The purpose of this note is to examine the consequences of this hypothesis for lepton colliders, in the past for LEP2 and in the future for ILC.

After a brief introduction to the 331 model, one will draw the consequences from the LHCb observation for the parameters of this model and the existing constraints, with particular emphasis on LEP2 and ATLAS+CMS. Two scenarios will be studied which can provide a reasonable explanation of the LHCb effects.

These two scenarios will be applied to the International Linear Collider (ILC) set up to see in which way it can allow for a decisive proof of this interpretation using lepton and quark final states. For the latter, ILC detectors can give a clean separation of c/b/t channels which provides a unique way to investigate the underlying structure.

Possible observations at LEP1 and GigaZ will be discussed which, altogether with high energy observations, allow a further elucidation of the underlying model.

After a brief discussion on prospects for ATLAS and CMS, I will conclude.

## II. 331 for pedestrians

The $SU(3)_c \otimes SU(3)_L \otimes U(1)_X$ model has been developed for 20 years[3,4]. It consists mainly in extending $SU(2)_L \otimes U(1)_Y$ to $SU(3)_L \otimes U(1)_X$ where $SU(3)_L$, in analogy to $SU(3)_c$, contains triplet representation made out of ordinary fermions plus new heavy fermions giving (ν, $\ell$, F) left-handed triplets for leptons and for quarks while right handed fermions belong to singlets. In analogy to the SM, one can write down the charge of the fermions in terms of the diagonal generators of the group:

$$Q = T_3 + \beta T_8 + X \mathbb{1}$$

$T_3$ and $T_8$ being the diagonal Gell-Mann matrices $T_3 = 1/2(1,-1,0)$ $T_8 = 1/2\sqrt{3}(1,1,1-2)$ and $\mathbb{1}$ is the unit matrix (1,1,1).
The value of β determines the charges of the exotic spectrum. To avoid exotic charges one usually only considers $\beta = \pm 1/\sqrt{3}$ and $\beta = \pm\sqrt{3}$.

A 331 anomaly free model requires, combining the 3 quark and 3 lepton families in an equal number of triplets and anti-triplets. This can only be fulfilled by having one quark family (3 triplets taking into account colors) in the same state as the leptons (3 triplets or anti-triplets) and the other 2 in opposite states (6 triplets or anti-triplets. This quark family can be any of the 3 quark families however the 'natural choice' seems to be the 3d family but this not necessarily the case.

From this first discussion one can understand that there are various choices which are allowed by the 331 model. It is also clear that this model relates the number of colors to the number of families which represents a major improvement with respect to the SM. With a non universal coupling to the 3 families, this model also allows non diagonal couplings at tree level which could play an essential role in strangeness and beauty phenomenology but with a CKM type suppression which insures minimal flavor violation. This is for instance the case for Z'->bs where the coupling is suppressed by Vts*Vtb.

Having chosen the 3d quark family, one has two possibilities: leptons and t/b are in a triplet or an anti-triplet. It can be shown that going from triplet to anti-triplet amounts to change the definition of β (β to – β ) and therefore does not matter. In the following I will choose the conventions from [5].

## III. What can already be said about a 331 Z'

### 3.1 LHCb

An anomaly has been observed in K*µµ data which can be interpreted in terms of the 331 model through the bs->Z'->µµ process under the following constraints:

- A large vector coupling for Z'->µµ with β=–√3 and Mz'=7 TeV which can cope with the Bs->µµ constraint
- A coupling for Z'->bs which goes like Vts*Vtb which can cope with the Bs-Bsbar mixing constraint

Reference [2] claims that these conditions can be met for B physics but ignoring the two following constraints:

- LEP2 measurements for the two fermion final states
- The running mass scale for s²w

The former is discussed below, while the latter can strongly affect the coupling constant of fermions for a heavy Z' for the solution elected in [2]. The origin of the problem can be easily understood by noting the Z' coupling constant g²X can be written as [5]:

$$g_X^2 = \frac{6g^2 s_W^2}{1-(1+\beta^2)s_W^2}$$ where g and s²w are the usual SM quantities (in [7] $g_X$

has a similar definition but w/o the factor 6)

For the solution β=-√3 chosen in [2] one sees that the denominator goes like 1-4s²w which is naturally small given that s²w=0.231 resulting in a large coupling constant. As already noted in [7]one cannot ignore the running of s²w which reaches 0.25 for a mass scale of ~4 TeV. This effect becomes dramatic if Mz'~7 TeV as in ref [2]. There are two possible ways out to cope with this problem:
1/ either renounce to the solution β=-√3 which means decreasing $g_X$ and lowering Mz' to get a significant effect at LHCb. For β=±1/√3 this would mean adopting a Z' mass which is already excluded by ATLAS/CMS Z' direct searches. Therefore ref [6] has used an intermediate value **β=-2/√3** which gives fractional charges. The corresponding limit Z' mass limit from ATLAS/CMS is **~3.2 TeV**.

Choosing the minimum allowed value for Mz' the LHCb effect gets reduced by 0.6 which seems acceptable given the present uncertainties.

2/ or assume that the running of s²w is reduced by the contribution of new particles[7]. This in turn means that these particles could be seen at LHC14.

For the present purpose, this means that $g_X$ is not a priori known for a given β but one also needs to know the running law governing s²w.

In any case one should realize that the LHCb effect only allows to estimate the following quantity:

$$\frac{g_X^2 V_{ts}^* V_{tb} V_e}{M_{Z'}^2}$$

where Ve and $g_X$ depend on our choice for β.

Having fixed β one can extract g²$_X$/Mz'² and this allows to test if the model passes the various constraints, in particular LEP2 and Bs mixing. Note that this also means that for β=-√3 one cannot determine unambiguously Mz' since this requires taking into account the running of s²w. This of course impacts on the predictions for ATLAS/CMS.

In the future, LHCb will keep the best sensitivity on the K*μμ mode with an excellent trigger and reconstruction efficiency but one can expect systematic effects to limit the significance of this signal.

### 3.2 LEP2

LEP2 data can already provide tight constraints for the solutions which are presently discussed to interpret the LHCb effect. The simple reason for this is the requirement of a strong vector coupling to lepton pairs. To vision simply what is going one can, to a good approximation, neglect the Z vector coupling (v~1-4s²w~0) and the Z' axial couplings therefore write the following vector amplitude and axial amplitudes for leptons:

$$V = e^2 + C^2 V_e^2 BW_{Z'}, \quad A = \frac{e^2}{16 s_W^2 c_W^2} BW_Z$$

where BW=s/(s-M²) is the Breit Wigner function and where C and Ve are given in the appendix for both models. From these expressions one derives the cross section ~V²+A² and finds that at LEP2 this cross section, normalized to the SM, goes like Rℓ=1+20.4BWz' for model 1 and Rℓ=1+1.87BWz' for model 2.

For the quarks one similarly derives Rq=1+9.8BW for model 1 and Rq=1+0.9BW for model 2.

From LEP2 (see [8]) one can use 2 measured ratios:

- Rμ+τ=0.9958±0.0116  the ratio of the measured leptonic cross section to the SM prediction combining the μ and τ channels
- Rq=1.0092±0.0076 the ratio of the measured quark cross section to the SM prediction combining u+d+s+c+d channels

Using above quantities one can set a 95% C.L. on Mz' for both models. The following table summarizes these mass limits.

| Model | Mz' from leptons | Mz' from quarks |
|---|---|---|
| 1 | >5.5    TeV | >8    TeV |
| 2 | >1.6    TeV | >2.5  TeV |

For the quark sector the improved sensitivity comes from a better experimental accuracy and from the positive fluctuation of the measurement which goes against the predicted negative interference.

From these evaluations one concludes that, for model 1, a Z' with a mass of order 8 TeV fulfills all criteria, still allowing a sufficient contribution to the LHCb effect but with the caveat due to the unknown running of s²w. This running of s²w can easily modify the mass limit from LEP2. For instance if we take s²w(Mz')=.215 as allowed in [7], one has $g_X$=2 hence the limit on Mz' drops to 6 TeV. Note that with the latter solution nothing would be changed for the quantitative contribution of the Z' to the LHCb effect since g²X/Mz'² would remain the same.

For model 2, the limit from ATLAS and CMS on the mass distribution is already reaching **3.2 TeV** as deduced from figure 2 from section 7 and I will therefore use this mass for model 2.

## IV.     Strategy with ILC

ILC operating at 500 GeV and with far better measurements accuracies and b/c tagging [9] than at LEP will allow to very precisely measure the fermion couplings to Z'. The figure below gives a precise estimate of the tagging purity/efficiency of the future ILC detectors.

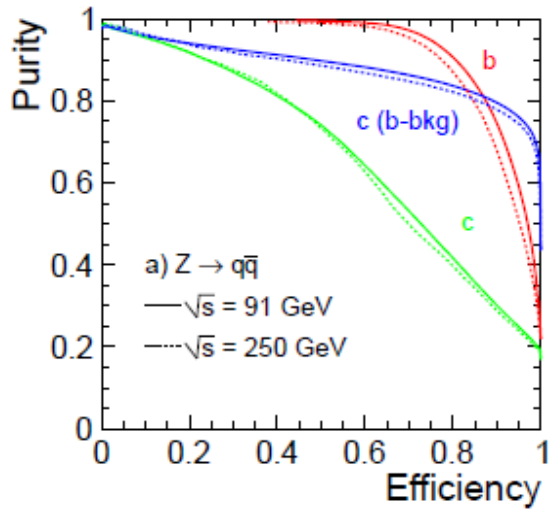

At ILC 500 GeV top pairs can be isolated with 70% efficiency and 90% purity [10]. b quark pairs can also be selected with  0.85²=70% efficiency and 4% contamination.
For what concerns c quarks, if one requires less than 10% contamination, the efficiency drops to 30%.

The sensitivity for the Z' mass goes like s/√$\mathcal{L}$ . For ILC at 500 GeV this is 60 times higher than at LEP2 which sets, for lepton couplings alone, the reach for Z' at 60 TeV for model 1 and 13 TeV for model 2, that is well beyond the sensitivity of LHC.
In addition, ILC can measure the couplings to the 3 'heavy quarks': t, b and c. By measuring the pattern of deviations on the cross sections for these 3 categories of quarks and for the leptons, one can unambiguously recognize the structure of 331 and also extract the value of β, which is quantized to avoid exotic charges for the new heavy fermions present in the model.

### 4.1 First model

|         | σ          | σF – σB    |
|---------|------------|------------|
| e+μ+τ   | -8.5±0.2%  | -4.8±0.4%  |
| d+s+u+c+b | -4.5±0.15% |          |
| Charm   | -7.6±0.4%  |            |
| Top     | -1.7 ±0.3% | -1.0±0.7%  |
| Bottom  | -4.9 ±0.4% | -4.75±0.7% |
| Down    | 2.8%       |            |

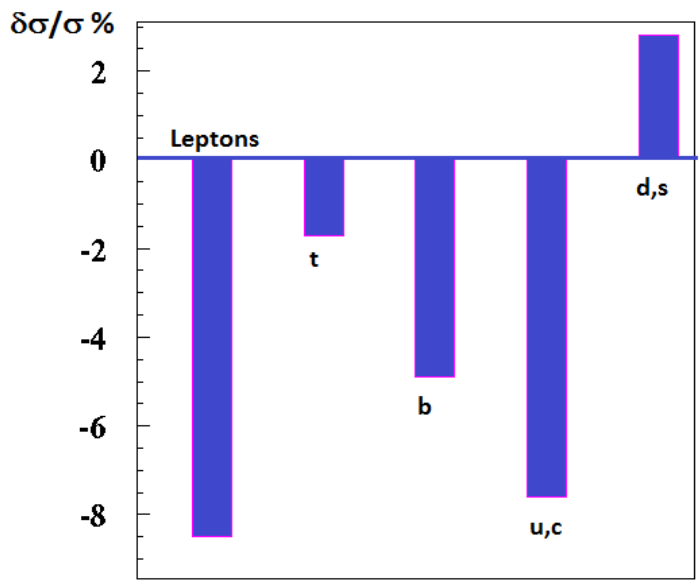

This model assumes Mz'=8 TeV and β=-√3.

The plot shows the pattern of expected relative variations of the cross sections with an estimated experimental error which only includes experimental errors, assuming an integrated luminosity of 500fb-1 taken at 500 GeV with a ±0.1% accuracy. The table also gives the relative variations for the differences of forward minus backward cross sections which can be precisely measured for leptons, b and top quarks.

### 4.2 Second model

|         | σ           | σF – σB   |
|---------|-------------|-----------|
| e+μ+τ   | -5.2±0.2%   | -3.0±0.4% |
| d+s+u+c+b | -2.75±0.15% |         |
| Charm   | -4.94±0.4%  |           |
| Top     | 0.13 ±0.25 %| 0.7±0.7%  |
| Bottom  | -4.25 ±0.4% | -4.2±0.7% |
| Down    | 3.1%        |           |

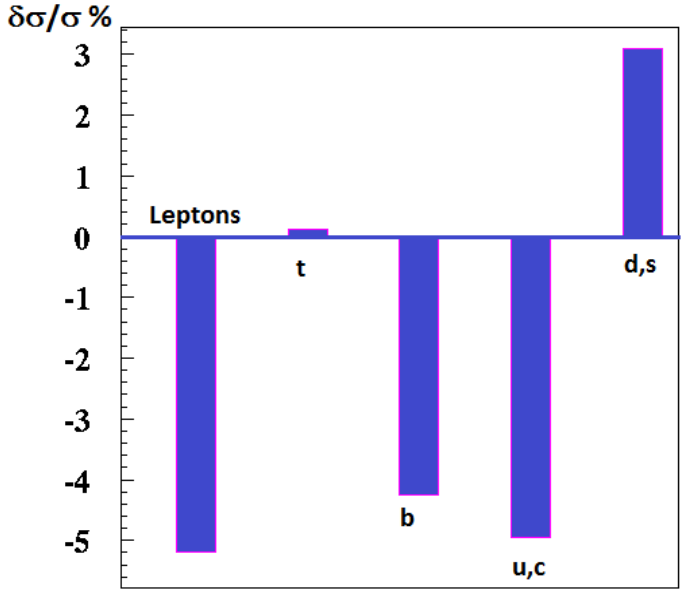

Table and plots are given for Mz'=3.2 TeV and =-2/√3 .

It is straightforward to determine β from these figures by, for instance using the ratio :

$$R_{b\mu} = \frac{\left(\frac{\delta\sigma}{\sigma}\right)_b}{\left(\frac{\delta\sigma}{\sigma}\right)_\mu} \sim K \frac{c_W^2 - x/3}{c_W^2 - 3x} \quad \text{where } x=\beta\sqrt{3}s^2w$$

which comes out very different for both models . This ratio depends on β and is independent of Mz' and $g_X$. For model 1 Rbµ=0.58±0.05 instead of 0.82 for model 2 which already gives almost 5σ discrimination.

Knowing β it is possible compute $g^2_X/M^2z'$ where $g_X$ is precisely known given β and therefore extract Mz' to better than a %. It is however fair to add that for this determination one encounters again the issue of the running of sw².

### 4.3 Axial couplings

With the solutions used to explain the LHCb anomaly, one expects an almost vanishing axial coupling of Z' to leptons. As discussed in appendix this feature can be tested by confronting σF – σB with σ providing a further test on the 331 model. Note that the very small effect seen on the top quark on these quantities comes from the vanishing vector coupling of Z' to top quarks which also is an important ingredient to test the two models.

## V. GigaZ and LEP1/SLC

For the measurement of s²w, LEP1/SLC results agree on average with the SM but SLC results based on the ALR measurement are more than 3 sd away from the most precise result of LEP1 based on the AFBb (see figure 4 in appendix). The latter effect can be explained by several extensions of the SM which either predict mixing between the quarks of the 3d family and new vectorlike quarks or a mixing of the Z boson with a Z' boson which couples preferentially to the 3d family. The 331 model can also provide for such an explanation with quark mixing.
The next issue is to ask the size of the Z-Z' mixing within 331 and what would be the impact on the LEP1/SLC measurement.

The mixing angle goes like $g_X/M_{z'}^2$ and can be sizeable under the two scenarios under consideration. In the 1st scenario, as discussed in appendix, the large value of $g_X$ produces a mixing angle below 10-3, but comparable to LEP/SLD combined accuracies (1.6 10-4). In the second scenario, with small Mz', the effect is even larger.

Below are given the mixing angles for the two models. These solutions are not unique since the 331 Higgs sector contains 3 vacuum expectations, one of them giving its mass to Z' the two others related to Z mass with a free parameters $\tan\alpha$ which is related to their ratio. In the following table θmix1 corresponds to cos(2α)=-1 and θmix2 to cos(2α)=1 and therefore define the two extremes.. These solutions give a negative shift on s²w which has the right magnitude to improve the agreement of the SLC measurement with respect to the SM prediction (-5 10-4 displacement as can be seen in figure 4 in appendix). Of course one can also argue that BSM physics allowed by 331 can contribute to the ρ parameter and modify the s²w prediction.

| Model | $g_X$ | Mz' TeV | θmix1 | θmix2 | ds²w1 | ds²w2 | ds²w LEP/SLD |
|---|---|---|---|---|---|---|---|
| 1 | 2.64 | 8 | -3.7 10-4 | -0.7 10-4 | -5 10-4 | -1.1 10-4 | ±1.6 10-4 |
| 2 | 1.08 | 3.2 | -7.5 10-4 | -1.5 10-4 | -12 10-4 | -2.3 10-4 | ±1.6 10-4 |

Recalling that the accuracy on s²w at GigaZ could reach **10-5** [11], this provides, in combination with the measurement at 500 GeV an access to tanα.

## VI. Z' solutions from other theories

A Z' from 331 has a very peculiar behavior as compared to other Z' models. In particular it has non universal coupling to up and top quarks which can be tested at ILC. This behavior is unusual in most of the Z' extensions of the SM.
RS models and little-H models also could produce a distinct coupling for top quarks either by privileged couplings to Z' bosons or by mixing with heavier fermions but they would not generate deviations on c quarks or leptons couplings.
A 331 model allows to fit the type of pattern shown previously by adjusting 2 parameters: β and $g^2_X/M^2_{z'}$. The former parameter can only take some quantized values which is also an unmistakable signature.
Finally GigaZ could provide a very substantial piece of information to confirm a 331 interpretation. As discussed below both scenarios corresponds to a Z' mass which fall within the reach of LHC. Also LHC can add major informations since it could produce new heavy fermions or charged bosons predicted by these models.

In particular charged bosons masses are simply related to the Z' mass by:

$$\frac{M_{Z'}^2}{M_V^2} = \frac{2g_X^2 c_W^2}{9g^2 s_W^2}$$

For scenario 1 this gives $M_V$=2.2 TeV and for scenario 2 $M_V$=2.14 TeV.

Concerning the fermions, the interpretation of the LEP1 anomaly for AFBb given in[7] predicts a heavy fermion with a mass of ~1 TeV which could be observed at LHC.

## VII. ATLAS and CMS

Z' production mainly occurs through fusion of a valence quark from the proton (2/3u, 1/3d) with a sea anti-quark. It is observed through a lepton pair and therefore the rate depends on its branching ratio to charged leptons. One can directly compare this rate to the so-called Z SSM rate with corresponds to a Z' with SM couplings, as given in the figures below.

For scenario 1 it turns out that the cross section is about 10 times larger than for a SM and assuming that there are no BSM decays has a branching ratio (BR) into leptons 3 times larger. From the figure 3 curve one concludes that even in the most optimistic scenario it would take the full power of HL-LHC to reach a discovery imit of 8 TeV. Note that the 331 curve given in figure 3 corresponds model 1 but with two caveats:

- It ignores the running of $s^2w$ which may change the cross section
- It ignores the contribution to the total width of heavy particles predicted by this model which would decrease the branching ratio into lepton pairs

For the second scenario the cross section is about the same as ZSSM and the BR into leptons could be up to 3 times larger which means that LHC14 should discover this Z' in its early running phase. This conclusion appears much less model dependent than for previous scenario since one can neglect the running of $s^2w$ to compute the rate.

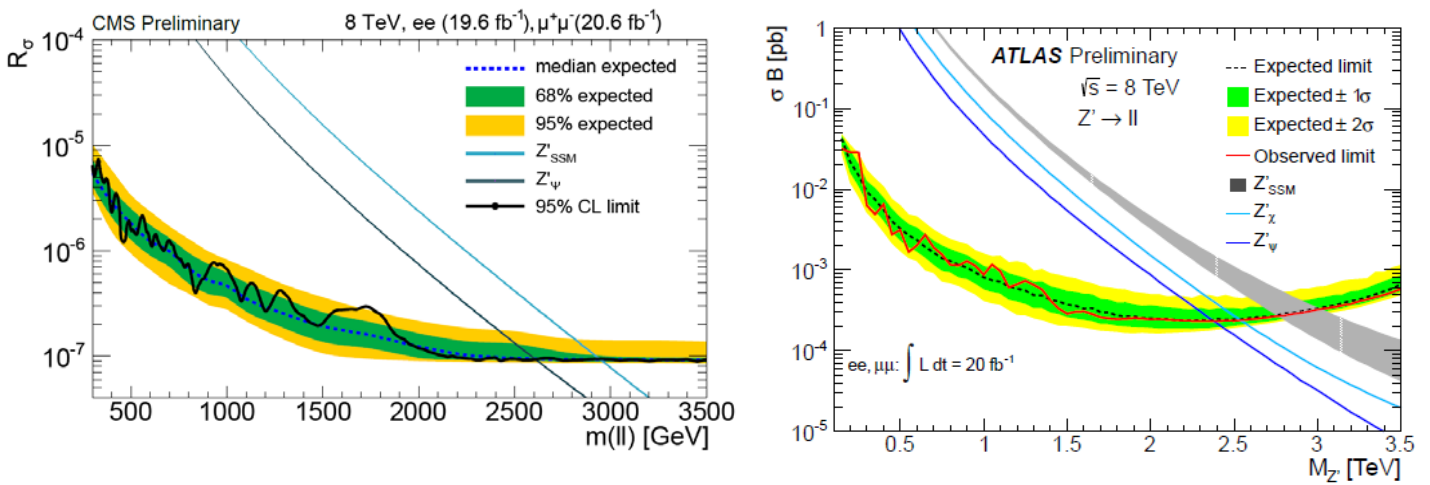

Figure 2; 95% C.L. exclusions from ATLAS and CMS [12,13] on σB(Z'->ℓ+ℓ−) where Z'SSM is a Z' with couplings identical to the Z.

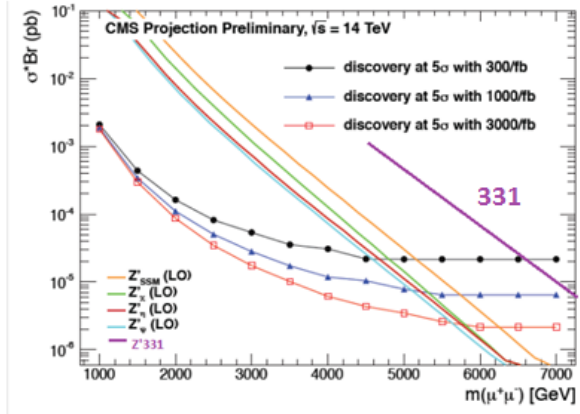

Figure 3: Extrapolated sensitivities of CMS on Z' searches. The line 331 corresponds to a 331 Z'.

As mentioned in the previous paragraph, LHC could add very important pieces of informations not only by observing directly the Z' but by observing some vector bosons or fermions predicted by this model.

# Conclusion

Given the very few indications of physics BSM, the evidence shown by the analysis of LHCb for K*µµ provides a precious example for evaluating the discovery potential of high energy colliders. Taking at face value the Z' exchange interpretation in the 331model, I have studied 2 examples of parameters compatible with the LHCb effect to investigate this potential.

In the first example, the reach of lepton colliders is well illustrated by the limit set by LEP2 on the Z' mass which surpasses by far LHC and reaches 8 TeV. It was then shown that ILC operating at 500 GeV with much higher luminosity and access to top quark couplings can unambiguously establish the 331 interpretation and precisely determine $\beta$, a key parameter of this model.
In the second example, where the Z' couplings are reduced, LHC sets the best lower limit on the Z' mass, ~3.2 TeV, I took this lowest mass which is still sufficient to interpret the effect seen at LHCb. At ILC I find that the same job can be fulfilled to precisely establish the 331 interpretation. Z-Z' mixing effects are also substantial and already at the limit of the LEP1 sensitivity. GigaZ could give a very precise measurement of the mixing angle and extend very significantly our understanding of the model . Furthermore it could confirm or disprove the AFBb anomaly observed at LEP1 and explore b quark mixing with the heavy quarks predicted by 331.
LHC will play an essential role in searching for the new heavy particles predicted by this model in particular the heavy quarks. In scenario 2 it would soon discover the Z'in the leptonic mode but would not provide enough information to establish the 331 interpretation.
In parallel LHCb and super-KEKb can develop various strategies to find additional proofs on the origin of this effect but this aspect does not belong to the present discussion.
In conclusion the 331 scenario offers a beautiful illustration of the complementary roles of high energy colliders. While ILC would unmistakably establish the underlying structure of 331, LHC could observe some of the heavy new objects predicted by this model already at 14 TeV.
Establishing such a scenario could set a goal for HL-LHC and/or would provide a strong motivation for a 100 TeV hadronic collider.

**Acknowledgements**

I warmly thank Andrezj Buras for very useful discussions.

# APPENDIX

## 1. Formulae

The following table uses [5] to compute the contributions from the 2 models chosen for 331.
As an example (model 1), for leptons one has the following formula for the vector amplitude:

$$V = e^2 + \frac{e^2}{4s_W^2 c_W^2}\left(-0.5 + 2s_W^2\right)^2 BW_Z + C_1^2\left(1 + 8s_W^2\right)^2 BW_{Z'}$$

where BW=s/(s-M²) are the Breit Wigner expressions for Z and Z'.

| β | -√3 | -2/√3 | SM |
|---|---|---|---|
| Coupling | C1 | C2 | e/2cwsw |
| Ve= 1-(1+3√3β)s²w <br> Ae=-1+(1-√3β)s²w | 1+8s²w <br> -1+4s²w | 1+5s²w <br> -1+3s²w | -0.5+2s²w <br> -0.5 |
| Vc =-1+(1+5√3β/3)s²w <br> Ac= 1-(1-√3β)s²w | -1-4s²w <br> 1-4s²w | -1-7/3s²w <br> 1-3s²w | 0.5-4/3s²w <br> 0.5 |
| Vt =Vc+2c²w <br> At=Ac-2c²w | Vc+2c²w <br> Ac-2c²w | Vc+2c²w <br> Ac-2c²w | 0.5-4/3s²w <br> 0.5 |
| Vb =Vd+2c²w <br> Ab=Ad -2c²w | Vd+2cw² <br> Ad-2c²w | Vd+2c²w <br> Ad-2c²w | -0.5+2/3s²w <br> -0.5 |
| Vd =-1+(1-√3β/3)s²w <br> Ad=1-(1+√3β)s²w | 1-2c²w <br> 1+2s²w | -1+5/3s²w <br> 1+s²w | -0.5+2/3s²w <br> -0.5 |
| Vν=1-(1+√3β)s²w=-Aν | 1+2sw² <br> -1-2s²w | 1+s²w <br> -1-s²w | 0.5 <br> 0.5 |

$$C1 = \frac{e}{4c_W s_W \sqrt{3 - 12s_W^2}} \quad C2 = \frac{e}{4c_W s_W \sqrt{3 - 7s_W^2}}$$

## 2. Axial couplings

Using cross sections alone one cannot separate the effect of BSM physics on V and A from the observed deviations.
The usual method is therefore to use the FB asymmetries recalling that these asymmetries are proportional to the product VA while the cross sections gives the combination S=A²+V².
It turns out that in the 331 hypothesis |dA/A|<<|dV/V|. This is true for the leptonic channel by construction since one needs dAe~0 to reproduce the LHCb data. For quarks dA/A is also very small due to initial state coupling. To observe A one needs to have a quark with a small vector coupling to Z' which is precisely the case of the top quark hence the very small variations observed both for the cross section and the asymmetry. In this sense the top channel offers a unique way of checking the 331 hypothesis.
One can write dFB/FB=dV/V+dA/A and dS/S=(2AdA+2VdV)/S. For the electron case A²<<V² and therefore one finds that dS/S~2dFB/FB which is almost true as seen from section 4 tables. This rule does not work for the bottom quark where the photon coupling is small and one has V~A but here also the Z' contribution to A is very small.

## 3. Masses & Z-Z' mixing

[5] defines:

$$g_X^2 = \frac{6g^2 s_W^2}{1-(1+\beta^2)s_W^2}$$ (while [7] does not have the factor 6).

Assuming a large vacuum expectation U:

$$M_{Z'}^2 \sim \frac{g^2 c_W^2 U^2}{3\left[1-(1+\beta^2)s^2 w\right]} = \frac{g_X^2 c_W^2 U^2}{18 s_W^2}$$

Since one measures $g_X^2/M_{Z'}^2$ by interference at high energy, $U^2$ can be unambiguously measured.
Charged bosons of type V and Y have masses given by $M^2 V = g^2 U^2/4$ hence:

$$\frac{M_{V,Y}^2}{M_{Z'}^2} = \frac{3\left[1-(1+\beta^2)s_W^2\right]}{4c_W^2}$$

The mixing angle is given by [7]:

$$\theta \sim \frac{2\sqrt{6} g g_X v^2 \left[3\beta s_W^2 + c_W^2 \cos(2\alpha)\right]}{18 s_W c_W^2 M_{Z'}^2}$$ where v~200 GeV and where tanα=vρ/vη is the ratio of

the two vacuum expectations contributing to the Z mass. Note that [7] uses anti-triplets for leptons which requires changing β into -β as commented in section II.
As numerical examples:

- Model 1 (no running of sw) $g_X$=2.64  Mz'=8 TeV  U=7 TeV  $M_V$=2.2 TeV
- Model 2 $g_X$=1.08  Mz'=3.2 TeV  U=6.9 TeV  $M_V$=2.1 TeV

As recently pointed out [14], such light vector bosons could provide a significant contribution for the g-2 anomaly.
The following table gives the resulting mixing angles and the corresponding deviations for s²w. The two extreme values correspond to the choices cos(2α)=±1.

| Model | $g_X$ | Mz' TeV | θmix1 | θmix2 | ds²w1 | ds²w2 | ds²w LEP/SLD |
|---|---|---|---|---|---|---|---|
| 1 | 2.64 | 8 | -3.7 10-4 | -0.7 10-4 | -5 10-4 | -1.1 10-4 | ±1.6 10-4 |
| 2 | 1.08 | 3.2 | -7.5 10-4 | -1.5 10-4 | -12 10-4 | -2.3 10-4 | ±1.6 10-4 |

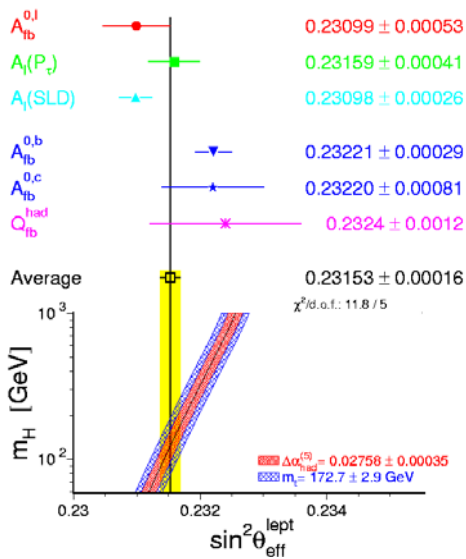

One sees that these mixing angles are large enough to be constrained at LEP1/SLC. The deviation on s²w with respect to the SM seen at SLC, ds²w~-5 10-4 (see figure 4) can be compensated in both models.
With this shift of s²w the disagreement with AFBb measured at LEP1 would be amplified but, as discussed in [7], quark-mixing can solve the problem.

**Figure 4 LEP1/SLC measurements of s²w**

13